\documentclass[aip,apl,superscriptaddress,amsmath,amssymb,floatfix,reprint]{revtex4-1}

\usepackage{graphicx}

\usepackage{float}

\usepackage{dcolumn}

\usepackage{bm}

\usepackage{units}

\usepackage[version=4]{mhchem}

\usepackage{amsmath}

\usepackage{color}

\usepackage[colorlinks=true, urlcolor=blue, linkcolor=black, citecolor=black]{hyperref}

\def\eV{\,\textrm{eV}}
\def\Bohr{\,\textrm{Bohr}}
\def\cm{\,\textrm{cm}}

\def\aa{\,\textrm{\AA}}
 
\def\VperA{\,\textrm{V}/\textrm{\AA}}

\def\Bi{\textrm{Bi}}
\def\Sb{\textrm{Sb}}
\def\Te{\textrm{Te}}
\def\Se{\textrm{Se}}

\begin{document}

\title{Gate field effects on the topological insulator $\Bi \Sb \Te \Se _ 2$ interface}

\author{Shuanglong Liu}
\affiliation{Department of Physics, University of Florida, Gainesville, Florida 32611, USA}
\affiliation{Quantum Theory Project, University of Florida, Gainesville, Florida 32611, USA}

\author{Yang Xu}
\altaffiliation{Current address: Applied and Engineering Physics, Cornell University, Ithaca, New York 14853, USA}
\affiliation{Department of Physics and Astronomy, Purdue University, West Lafayette, Indiana, 47907, USA}

\author{Yun-Peng Wang}
\altaffiliation{Current address: School of Physical Science and Electronics, Central South University, Changsha, Hunan, 410012, China}
\affiliation{Department of Physics, University of Florida, Gainesville, Florida 32611, USA}
\affiliation{Quantum Theory Project, University of Florida, Gainesville, Florida 32611, USA}

\author{Yong P. Chen}
\affiliation{Department of Physics and Astronomy, Purdue University, West Lafayette, Indiana, 47907, USA}
\affiliation{School of Electrical and Computer Engineering, Purdue University, West Lafayette, Indiana, 47907, USA}
\affiliation{Birck Nanotechnology Center, Purdue University, West Lafayette, Indiana, 47907, USA}
\affiliation{Purdue Quantum Science and Engineering Institute, Purdue University, West Lafayette, Indiana, 47907, USA} 

\author{James N. Fry}
\affiliation{Department of Physics, University of Florida, Gainesville, Florida 32611, USA}

\author{Hai-Ping Cheng}
\email{hping@ufl.edu} 
\affiliation{Department of Physics, University of Florida, Gainesville, Florida 32611, USA}
\affiliation{Quantum Theory Project, University of Florida, Gainesville, Florida 32611, USA}
\affiliation{Center for Molecular Magnetic Quantum Materials, University of Florida, Gainesville, Florida 32611, USA}

\date{\today}

\begin{abstract}

Interfaces between two topological insulators are of fundamental interest 
in condensed matter physics.
Inspired by experimental efforts, we study interfacial processes between 
two slabs of $\Bi \Sb \Te \Se_2$ (BSTS) via first principles calculations. 
Topological surface states are absent for the BSTS 
interface at its equilibrium separation, but our calculations show that 
they appear if the inter-slab distance is greater than $6 \aa $. 
More importantly, we find that topological interface states 
can be preserved by inserting two or more layers of 
hexagonal boron nitride between the two BSTS slabs. 
In experiments, the electric current tunneling through the interface 
is insensitive to back gate voltage when the bias voltage is small.
Using a first-principles based method that allows us to simulate 
gate field, we show that at low bias the extra charge induced by a gate voltage 
resides on the surface that is closest to the gate electrode, 
leaving the interface almost undoped. 
This provides clues to understand the origin of the observed insensitivity of 
transport properties to back voltage at low bias.
Our study resolves a few questions raised in experiment, 
which does not yet offer a clear correlation between microscopic physics 
and transport data. 
We provide a road map for the design of vertical tunneling 
junctions involving the interface between two topological insulators. 

\end{abstract}

\maketitle


Topological surface states (TSS) of a three dimensional topological insulator (TI) have drawn 
much research attention due to their robustness, linear dispersion, and spin-momentum 
locking, which allow potential applications in low-energy-consumption electronics and 
spintronics~\cite{Hsieh2012, Chen2012}.
A large category of electronic/spintronic devices involve an interface between a TI and 
another material, which could be a normal insulator, metal, magnet, superconductor, or 
even molecules~\cite{Gehring2012, Berntsen2013, Yoshimi2014, Seibel2012, 
NKim2013, Scholz2012, Shoman2015, JLi2012, MLi2015, 
LFu2006, Jauregui2018, JZhang2016,  Jakobs2015}.
It is thus necessary to understand how TSS are affected by proximity effects, 
especially their preservation~\cite{Gehring2012, Berntsen2013, Yoshimi2014, Seibel2012, 
NKim2013, Scholz2012, Shoman2015} or 
passivation~\cite{Berntsen2013, JZhang2016,  Jakobs2015}. 
A gate field is often applied to control the electron transport properties of 
an electronic device. 
A single gate induces charge doping, and a dual gate configuration can further 
create a vertical electric field. 
A gate voltage provides a knob for tuning topological surface/interface 
states~\cite{Steinberg2010, Sulaev2015, YXu2015}.
Nevertheless, theoretical work has been mostly limited to model~\cite{JLi2009, Yokoyama2010} 
or conventional electronic structure calculations.~\cite{Min2007, Houssa2016}

$\Bi _x \Sb _{2-x} \Te _y \Se _{3-y}$ (BSTS) has been reported to be topological insulator 
with high bulk resistivity and robust surface states 
for certain $x$ and $y$ values~\cite{AATaskin2011, TArakane2012, YXu2014, HLohani2017}.
In this work, we study the interface between two $\Bi  \Sb \Te \Se _2$ slabs, 
with and without spacers in between.
We construct a theoretical
approach that can simulate gate effects via a first-principles treatment.
The major results reported here are from theoretical investigations 
which are motivated by the experimental data.
%
Experimentally, two BSTS flakes are prepared separately before stacked together to 
form a vertical tunneling junction. 
It is observed that the electric current is not sensitive to a back gate voltage when the bias voltage 
between the two BSTS slabs is small (see Fig.~S1 in the supplemental material).
Based on these observations, theoretical investigations focus on the three questions --  
First, \textit{do topological interface states exist at the BSTS interface?} 
Second, \textit{how do topological interface states respond to 
a gate field when they are present?}
and, Third, \textit{how can we turn BSTS into a quantum system that is useful for future electronics?} 
This work offers a picture at the electron level, deepens our understanding of TI interfaces 
and demonstrates a way to utilize quantum TSS, and finally provides an idea for designing 
TI-based vertical tunneling devices.

\begin{figure}[ht!]
\centering
\includegraphics[width=0.28\textwidth]{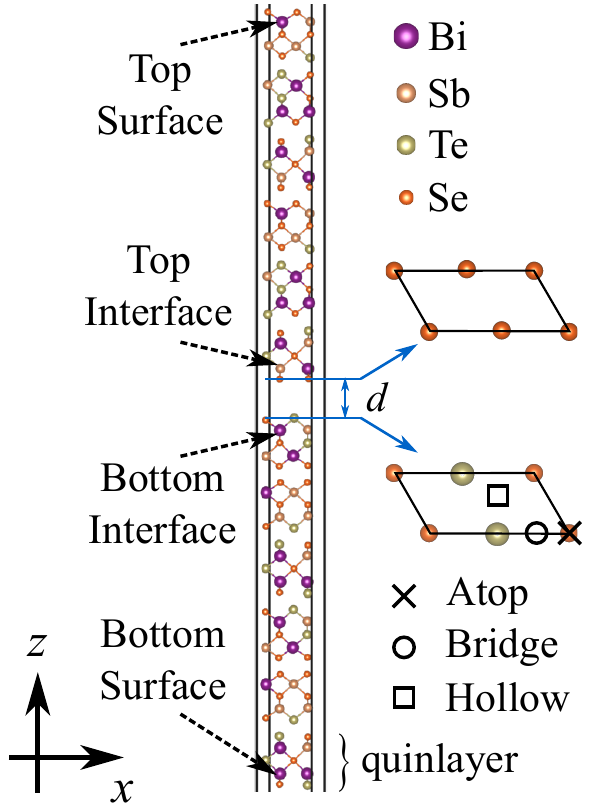}
\caption{\label{fig:pos}
Atomic configuration of the BSTS interface, consisting of two BSTS slabs. 
The two slabs are separated by a distance $d$ and each slab contains six quinlayers. 
The violet, ocher, olive and orange balls represent Bi, Sb, Te and Se respectively.
There are four ``surfaces'' in the system, denoted as the top surface, the bottom surface, 
the top interface, and the bottom interface, as indicated by the dashed black arrows. 
The interfacial atomic layers are zoomed in and shown by the side in a top view. 
Three stacking sites are marked in the interfacial atomic layer from the lower BSTS slab.  
The vertical solid lines are the boundaries of the unit cell. 
The interface is periodic in the $x$ and $y$ directions and perpendicular to the $z$ direction. 
} 
\end{figure}

In order to answer whether the interface hosts topological surface states, we first 
generate a special quasirandom structure of the BSTS alloy, 
that approximates the true disordered state with a periodic supercell~\cite{deWalle2013}, 
using the ``\texttt{mcsqs}'' 
code of the Alloy Theoretic Automated Toolkit~\cite{deWalle2013, deWalle2002}. 
The topological invariants of our quasirandom BSTS supercell are calculated to be
$(1;0,0,0)$ with the aid of the Z2Pack package.~\cite{Gresch2017,Soluyanov2011}
As such, our model BSTS represents a strong topological insulator. 
%
%
%
Then, we build a BSTS interface based on the special quasirandom structure and 
visualize it using the Visualization for Electronic and Structural Analysis (VESTA) 
software~\cite{Momma2011}. 
As shown in Fig.~\ref{fig:pos}, the interface is made of two BSTS slabs 
that are separated by distance $d\/$, where each slab contains six quinlayers, 
periodic in the $x$-$y$ plane perpendicular to the $z$-direction. 
The in-plane lattice constants are $a=8.390 \aa$ and $b=4.191 \aa$, 
with $\gamma = 120^\circ$. 
At least $15 \aa$ of vacuum is added along the $z$-direction to avoid 
interaction between periodic images of the same system. 
For a single BSTS slab with six quinlayers, 
we test that there are localized surface states with spin-momentum locking 
that disappear if spin-orbit coupling is switched off, 
evidence that the special random structure represents 
a topological insulator. 
Our calculations are based on density functional theory (DFT)~\cite{HK1964, KS1965} 
as implemented in the Vienna Ab initio simulation package (VASP).~\cite{KH1993, GKresse1999}  
In VASP calculations, we adopt at least a $450 \eV$ energy cutoff 
for plane waves, 
the PAW pseudopotential~\cite{PEBlochl1994, GKresse1999}, the optB86b vdW-DF 
energy functional for including the van der Waals interaction~\cite{
MDion2004, GRP2009, JKlimes2010}, 
a $5 \times 9 \times 1$ Monkhorst-Pack $k$-point mesh, 
a $1\times10^{-6}\eV$ energy tolerance for self-consistency, 
and a $0.01 \eV/\textrm{\AA}$ force tolerance for ionic relaxation. 

\begin{figure}[ht!]
\centering
\includegraphics[width=0.48\textwidth]{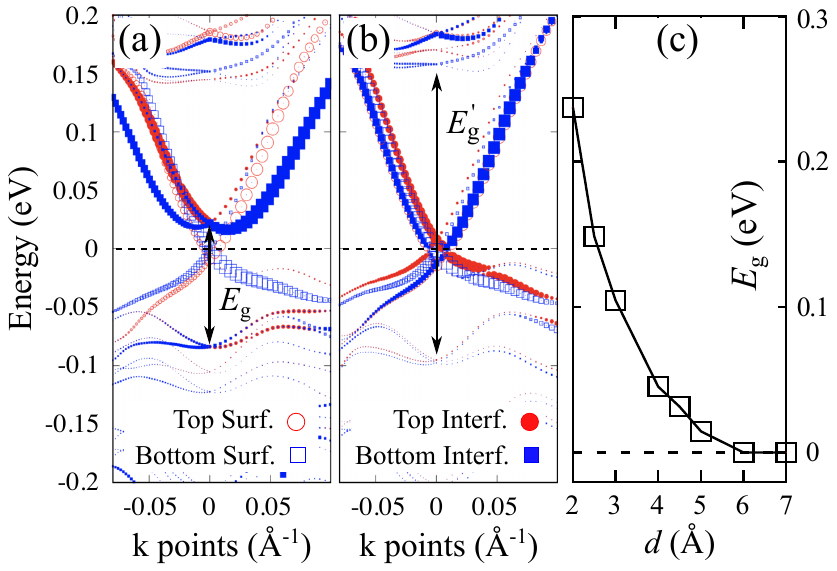}
\caption{\label{fig:bands} 
Band structure for a BSTS interface with an inter-slab distance 
of (a) $d = 3.0 \aa$ and (b) $d = 6.0 \aa$. 
The Fermi energy is set to zero.
Each dot represents a Kohn-Sham state, with size proportional to 
the projected density of states to the first quinlayer of a surface.
Red empty (filled) circles are for the top surface (interface) states; 
and blue empty (filled) squares are for the bottom surface (interface) states. 
If a dot is bigger, the state is more localized on the corresponding surface/interface. 
$E_g$ ($E_g^{\prime}$) is the energy gap for the interface (bulk) states 
at the $\Gamma$ point. 
(c) $E_g$ versus $d\/$ for the BSTS interface. 
} 
\end{figure}

Now, we examine the relation between interface states and the inter-slab distance $d$. 
%
%
Figs.~\ref{fig:bands}(a) and \ref{fig:bands}(b) show the band structure 
of the BSTS interface with an inter-slab distance of $3.0 \aa$ 
and $6.0 \aa$ respectively. 
For both $d=3.0 \aa$ and $d=6.0 \aa$, the top (bottom) surface band 
forms a Dirac cone around the Fermi energy. 
In contrast, for $d=3.0 \aa$, the top (bottom) interface band does not form 
a Dirac cone, and an energy gap of $E_g \approx 0.105 \eV$ opens at the 
$\Gamma$ point.
This energy gap however closes when $d=6.0 \aa$, 
due to weak interaction between the two slabs, and 
accordingly the Dirac cone for the top (bottom) interface is recovered. 
Therefore, the presence of topological interface states strongly depends 
on the inter-stab distance $d$. 
Fig.~\ref{fig:bands}(c) shows that $E_g$ decreases with $d$ 
and reaches zero at around $6 \aa$. 
Note that there are three typical alignments of stacking between 
the two BSTS slabs, namely the atop, the bridge and the hollow. 
The above discussion is for the BSTS interface with atop stacking; however
it remains valid qualitatively for the other two stackings. 
The energetically favorable $d$ for the atop (the bridge, the hollow) stacking
is about $3.6$ ($2.9$, $2.6$) {\AA} and a Dirac cone is not formed. 
The hollow site stacking is the most stable configuration, 
about $0.08 \eV$ and $0.30 \eV$ lower than the bridge stacking 
and the atop stacking respectively.

\begin{figure}[ht!] 
\centering
\includegraphics[width=0.42\textwidth]{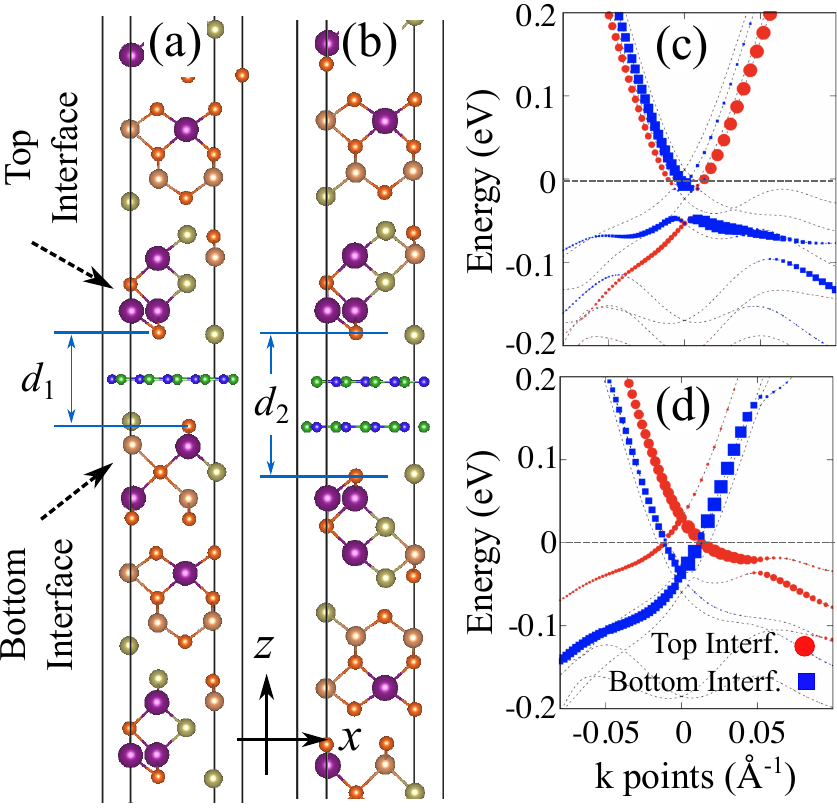}
\caption{\label{fig:posbn} 
Atomic configuration of the BSTS interface are shown 
with (a) monolayer $h$-BN and (b) bilayer $h$-BN. 
The optimized inter-slab distance is $d_1 = 7.04 \aa$ for panel (a) 
and $d_2 = 10.28 \aa$ for panel (b), as measured by the difference 
in the $z$-coordinate of the opposing $\Se$ atoms. 
%
%
Only the atoms close to the interface are shown, although there are six 
quinlayers in each BSTS slab. 
The vertical lines are the boundaries of the unit cell. 
Band structure for the BSTS interface is shown for 
(c) monolayer $h$-BN and (d) bilayer $h$-BN. 
The red circles (blue squares) are for the top (bottom) interface states. 
If a red circle (blue square) is bigger, the corresponding state is more localized 
at the top (bottom) interface. 
The dashed lines show all energy bands within the energy range 
$[-0.2:0.2] \eV$. 
The Fermi energy is set to zero, as indicated by the horizontal grey line. 
} 
\end{figure}

In order to recover TSS at the interface, we considered inserting $h$-BN 
between the two BSTS slabs, since $h$-BN is a normal insulator~\cite{WAuwarter2004} 
and TSS are known to exist at the interface between a topological insulator 
and a normal insulator~\cite{Gehring2012} (experimentally, it might be easier to
prepare BSTS with a surface covered by a $h$-BN monolayer before stacking 
two pieces together).
In the following, we show that monolayer $h$-BN is not sufficient, 
and two layers of $h$-BN are required. 
Fig.~\ref{fig:posbn}(a) [\ref{fig:posbn}(b)] shows the atomic configuration 
of the BSTS interface with monolayer (bilayer) $h$-BN, 
denoted as BSTS/$1$BN/BSTS (BSTS/$2$BN/BSTS). 
%
%
The band structures for BSTS/$1$BN/BSTS and BSTS/$2$BN/BSTS are shown in 
Figs.~\ref{fig:posbn}(c) and  \ref{fig:posbn}(d). 
The red circles (blue squares) are for the top (bottom) interface and again the size of
a circle/square indicates the localization of the corresponding state. 
When there is only one layer of $h$-BN, a band gap opens at the $\Gamma$ point 
for both the top and the bottom interface states around the Fermi energy. 
In other words, Dirac cone and TSS are absent at the interface. 
In contrast, when there are two layers of $h$-BN, the energy gap closes and 
two Dirac cones are formed around Fermi energy, one for the top interface states
and the other for the bottom interface states. 
These results are for hollow stacking between BSTS and BN; however they remain
valid for atop stacking and bridge stacking.
We observe that the top (bottom) interface is slightly 
hole- (electron-) doped. 
This charge transfer arises from the asymmetry in the atomic configuration of the interface. 
If we construct an inversion symmetric interface, then the two Dirac cones 
indeed become identical and the Dirac points are located at the Fermi level. 
%
%
Note that the top and the bottom surface states still exist, 
although they are not emphasized in Fig.~\ref{fig:posbn}. 
%

\begin{figure}[ht!]
\centering
\includegraphics[width=0.48\textwidth]{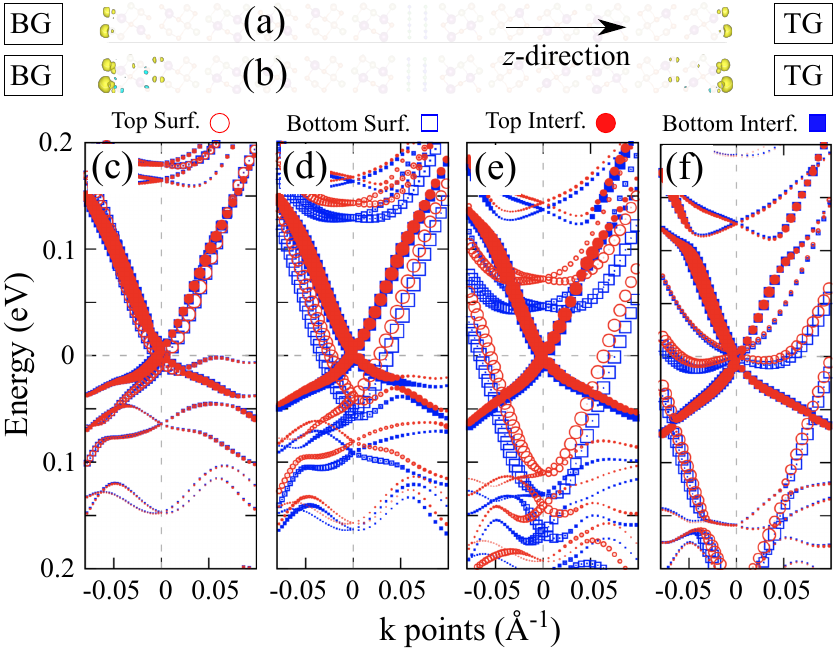}
\caption{ \label{fig:dope} 
Doping effects on the BSTS interface with bilayer $h$-BN. 
(a) Electron density difference $\rho(Q=-0.05, E=0) - \rho(Q=0, E=0)$, 
where $\rho(Q, E)$ is the electron density for the system with net charge $Q$ 
(measured in unit charges) per cell under electric field $E$ ($\VperA$). 
BG denotes the back gate electrode and TG the top gate electrode. 
(b) Electron density difference $\rho(Q=-0.10, E=0) - \rho(Q=0, E=0)$. 
The isosurface threshold for both (a) and (b) is $5 \times 10^{-5} \Bohr^{-3}$. 
Yellow (cyan) represents positive (negative) charge density, 
gaining (losing) electrons. 
(c)--(f): Energy bands of the system with a net charge of (c) $Q=0$, 
(d) $Q=-0.03$, (e) $Q=-0.05$, and (f) $Q=-0.10$, per cell. 
The electric field $E$ is zero for panels (c)--(f). 
The size of an empty/filled red circle is proportional to the local density of states of 
the top surface/interface. 
The size of an empty/filled blue square is proportional to the local density of states of 
the bottom surface/interface. 
} 
\end{figure}

So far, we have answered the question of whether or not TSS exist at the BSTS interface. 
Now we turn to the second question: 
How do topological interface states respond to a gate field? 
%
In order to answer this question, we take the BSTS/$2$BN/BSTS interface as 
an example and place it between two gate electrodes, as shown in Fig.~\ref{fig:dope}. 
Each gate electrode has a constant Hartree potential and the two gate electrodes could 
have different Hartree potentials, forming a non-periodic boundary condition.
We simulate the gate field effect using the Effective Screening Medium (ESM) method 
as implemented in the SIESTA package~\cite{ESM,SIESTA}, where 
the non-periodic boundary condition for the Hartree potential is dealt with Green's functions. 
A vacuum layer of about $1.2$ nm thick is inserted between the BSTS slab and the 
top/bottom gate electrode as a dielectric layer. 
We adopt norm-conserving relativistic pseudopotentials as generated via the 
Troullier-Martins scheme of the ``\textsc{atom}" code~\cite{NTroullier1991, ATOM}.  
The pseudopotentials for $\Bi$, $\Se$ and $\Te$ atoms are created by 
Rivero \textit{et al.}~\cite{PRivero2015} and the rest ones are created by the authors.
The localized basis set is optimized in order to obtain a reasonable band structure
compared with the results of VASP. 
We use the PBE exchange correlation energy functional.
This does not present a problem because the atomic structure is fixed and 
van der Waals interaction does not change the band structure much. 
A mesh cutoff of $150 \, \mathrm{Ry} $ is used to sample in real space.  



Two gate electrodes permit not only charge doping but also an electric field 
perpendicular to the interface. 
Here \textit{electric field} means the average electric field 
between the two gate electrodes, 
\begin{equation}
\label{eq:efield}
E = ( V_H^{\textrm{TG}} - V_H^{\textrm{BG}} )/L, 
\end{equation}
where $V_H^{\textrm{TG}} - V_H^{\textrm{BG}}$ is the Hartree 
potential difference between the top and bottom gate electrodes 
and $L$ is the distance between the two gate electrodes. 
The \textit{electric field} $E$ in the caption of Fig.~\ref{fig:dope} 
also means that in Eq.~(\ref{eq:efield}). 
Fig.~\ref{fig:dope}(a) and \ref{fig:dope}(b) show the charge redistribution 
across the BSTS interface when it is doped with extra electrons~\cite{note1} 
and subject to zero electric field. 
In Fig.~\ref{fig:dope}(a), the BSTS interface with bilayer $h$-BN is 
doped with $Q = -0.05 $ unit charges per unit cell.
The corresponding carrier density $\sigma = Q/S$ is about 
$-1.64 \times 10^{13} \cm^{-2}$, where $S$ is the surface area of a unit cell. 
As seen from the figure, the extra electrons are mainly located at the top and the bottom 
surfaces, while the inner part of the system is hardly doped. 
For a single gate configuration, the extra charge goes mainly to 
the surface that is closest to the gate electrode.
As such, the local electronic structure at the interface does not change much, 
and it might be a reason for the insensitivity of the electric current to 
the gate voltage at small bias voltage. 
In Fig.~\ref{fig:dope}(b), the net charge per cell is 
$Q = -0.1 $. 
In this case, compared with Fig.~\ref{fig:dope}(a) the extra electrons 
spread more into the inner part of the heterostructure.
In order to understand this, we plot the band structures of the BSTS 
interface with different net charges in Figs.~\ref{fig:dope}(c)--(f). 
%
As shown In Fig.~\ref{fig:dope}(c), 
without charge doping there are four Dirac cones around the Fermi energy 
for, respectively, the top surface, the top interface, the bottom interface, 
and the bottom surface, 
In Fig.~\ref{fig:dope}(d), the system is doped with $0.03$ electrons ($Q=-0.03$) 
per unit cell, and these electrons fill into the two Dirac cones for the the top and 
the bottom surfaces, leaving the remaining two Dirac cones undoped. 
This is consistent with the charge redistribution shown in Fig.~\ref{fig:dope}(a). 
It can also be seen from Fig.~\ref{fig:dope}(d) that the bulk energy bands 
above the Fermi energy move downwards. 
At yet higher doping levels such as  $Q=-0.05$ in Fig.~\ref{fig:dope}(e), 
the surface energy bands are filled with more electrons and the bulk 
energy bands move even closer to the Fermi energy. 
Eventually the bulk energy bands reach the Fermi energy and thus are also doped, 
as seen in Fig.~\ref{fig:dope}(f) where $Q=-0.10$ . 
Under certain doping levels, the charge distribution of the BSTS interface 
with bilayer $h$-BN can be further tuned by the electric field $E$ 
between the two electrodes. 
In the case of $Q=-0.05$, a small electric field along the $z$-direction 
moves electrons from the top surface to the bottom surface, and 
the inner part of the system is not doped until $E$ is greater than 
$0.01 \VperA$. 
This may be related to the increased sensitivity of the electric current 
to the gate voltage at higher bias voltages, as shown in Fig.~S1 
(see supplemental information).

In conclusion, we have a clear microscopic picture of the interface between two slabs of the 
topological insulator $\Bi \Sb \Te \Se _2$. 
We find that topological interface states are absent unless 
the inter-slab distance is greater than $6 \aa$; 
they can however be preserved by two or more layers of 
$h$-BN that serves as a spacer between the two BSTS systems, which is verified by 
our calculations and is the answer to the third question imposed at the beginning 
of this paper.
We undercover the mechanism underlying physical processes in small and large doping levels and
provide clues to understand electron transport characteristics 
in the BSTS vertical tunneling junction configuration.
The combination of a $h$-BN spacer and single/double gating is a promising way to protect TSS and
modify interfacial electronic processes.

See the supplemental information for 
electron transport measurement of a BSTS vertical tunneling junction, 
energy bands and hybrid Wannier charge centers of the quasirandom BSTS, 
results for different stackings between a BSTS slab and another BSTS slab or a BN layer, 
charge distribution for a single gate configuration, 
effect of electric field on the BSTS interface, 
parameters for basis set and pseudopotentials etc.


This work is supported by the US Department of Energy (DOE), 
Office of Basic Energy Sciences (BES), under Contract No. DE-FG02-02ER45995. 
Y. P. C. acknowledges partial support from National Science Foundation (NSF) 
under Grant No. EFMA-1641101. 
Computations were done using the utilities of the National Energy Research 
Scientific Computing Center and University of Florida Research Computing.



\begin{thebibliography}{11}
\bibitem{note1}
Since the TSS below the Fermi energy are not very accurate in SIESTA compared with VASP, we will consider only electron doping in the following.
\end{thebibliography}
\end{document}